\documentclass[conference]{IEEEtran}
\pdfoutput=1
\usepackage{amssymb}
\usepackage{booktabs}
\usepackage[noadjust]{cite}
\usepackage{enumitem}
\usepackage[pdftex]{graphicx}
\usepackage[utf8]{inputenc}
\usepackage[framemethod=tikz]{mdframed}
\usepackage{tikz}
\usepackage{tipa}
\usepackage{xtab}
\usepackage{todonotes}
\usepackage{url}
\usepackage{relsize}

\title{Do as I Do, Not as I Say:  Do Contribution Guidelines Match the GitHub Contribution Process?}

\author{
	\IEEEauthorblockN{Omar Elazhary\IEEEauthorrefmark{1}, Margaret-Anne Storey\IEEEauthorrefmark{1}, Neil Ernst\IEEEauthorrefmark{1} and Andy Zaidman\IEEEauthorrefmark{2}
	}
	\IEEEauthorblockA{
		\IEEEauthorrefmark{1}
		University of Victoria, 
		omazhary@uvic.ca, mstorey@uvic.ca, nernst@uvic.ca
	}
	\IEEEauthorblockA{
		\IEEEauthorrefmark{2}
		Delft University of Technology, 
		a.e.zaidman@tudelft.nl
	}
}

\begin{document}

\setlist{nolistsep}

\maketitle

\begin{abstract}
    Developer contribution guidelines are used in social coding sites like GitHub to explain and shape the process a project expects contributors to follow.
    They set standards for all participants and ``save time and hassle caused by improperly created pull requests or issues that have to be rejected and re-submitted'' (GitHub).
    Yet, we lack a systematic understanding of the content of a typical contribution guideline, as well as the extent to which these guidelines are followed in practice.
    Additionally, understanding how guidelines may impact projects that use Continuous Integration as part of the contribution process is of particular interest.
    To address this knowledge gap, we conducted a mixed-methods study of 53 GitHub projects with explicit contribution guidelines and coded the guidelines to extract key themes.
    We then created a process model using GitHub activity data (e.g., commit, new issue, new pull request) to compare the actual activity with the prescribed contribution guidelines.
    We show that approximately 68\% of these projects diverge significantly from the expected process.
\end{abstract}


\section{Introduction}
\label{sec:intro}

Open source software projects are the epitome of collaboration.
They represent the amalgamation of the work and effort of hundreds or thousands of developers coming together to achieve a single purpose: to create an application that fulfills user need.
However, there is a point where such a large workforce becomes too difficult to manage.
While public-facing, open source projects encourage contributions in general, some evidence by Gousios et al. \cite{gousios2016work} suggests maintainers can become overwhelmed with new contributions.
These contributions may frequently duplicate one another or repeat discussions in which the maintainer stated that a particular design choice was not going to be changed.
For some maintainers, the workload is simply too much.

Social coding sites like GitHub have started offering solutions, such as contribution guidelines and continuous integration (CI) tools, to get core developers and contributors on the same page and help unify expectations. Contribution guidelines and CI tools often go hand in hand. Contribution guidelines are textual documentation files that document the contribution expectations of project maintainers. 
In fact, GitHub considers contribution guidelines a prerequisite on an open source project's pre-launch checklist \cite{github2019guides} and provides a step-by-step tutorial on how to create such guidelines \cite{github2019contributing}.
Additionally, GitHub checks and refers contributors to the guidelines when they make a contribution \cite{barnes2012guidelines}.
As mentioned by Steinmacher et al. \cite{steinmacher2015systematic}, this form of documentation helps alleviate some barriers for new contributors.

On the more technical side of things, CI tools offer a way for developers to pool together their testing practices and evaluation criteria when it comes to assessing contributions~\cite{beller2017oops}.
Depending on how the tool is configured, it will run tests on submitted contributions and make those results available to anyone reviewing them.
The use of CI increases the efficiency of the contribution process and contributes to the quality of the code~\cite{vasilescu2015quality}. While previous research by Kobayakawa and Yoshida \cite{kobayakawa2017github} and another study by Prana et al. \cite{prana2018categorizing} attempted to explore the contents of contribution guideline documentation, they only focused on the contents of \emph{README} files.
They did not, however, consider if these guidelines match the reality of the development process. 
We do consider if these guidelines match the contribution process, but focus on projects that use CI, as we expect the contribution guidelines may be more prescriptive for those projects. 
The research questions we aimed to answer are as follows:
\vspace{-1mm}
\begin{mdframed}
   \begin{enumerate}[label=\textbf{RQ\arabic*:},leftmargin=*]
       \item What is the content of contribution guidelines for projects on GitHub?
       \item Do projects that use CI tools mention these tools in their contribution guidelines?
       \item To what extent do the actual processes in projects that use CI tools match their guidelines?
   \end{enumerate}
\end{mdframed}
\vspace{-3mm}
We present preliminary evidence that the contribution process prescribed in the contribution guidelines differs from what we observe in reality.
We also demonstrate that CI tools are only discussed as testing mechanisms and generally do not have documentation describing how they function or what they test.
\section{Background}
\label{sec:bkgnd}

We present related research on contribution guidelines and continuous integration tools.

\subsection{GitHub Contribution Guidelines}

As mentioned in Section \ref{sec:intro}, contribution guidelines are a way for core developers to communicate their expectations, both in terms of contribution criteria and processes, to developers who wish to contribute to a software project.
As such, contribution guidelines are considered an important addition to a project's overall documentation and many view a project as 
\emph{incomplete} without them \cite{github2019guides}.

Additionally, contribution guidelines offer a way for newcomers to orient themselves and learn the project's building blocks, processes, and other conventions laid down by developers.
In fact, Steinmacher et al. \cite{steinmacher2015systematic} illustrate that the lack of such documentation poses a barrier to entry for developers who wish to contribute to open source projects.

In an effort to bring the importance of contribution guidelines to the attention of developers, GitHub uses a reminder when creating an empty repository that allows developers to create a \emph{README.md} file with a single click.
They explicitly mention: ``We recommend every repository include a \emph{README}, $LICENSE$, and $.gitignore$.''
And while \emph{README} files do not necessarily give the impression of something that contains contribution guidelines, Prana et al. \cite{prana2018categorizing} demonstrate that they usually do.
Additionally, as mentioned previously, GitHub \emph{actively} reminds contributors of the existence of contribution guidelines and suggests they be inspected before making a contribution \cite{barnes2012guidelines}.

Prana et al. \cite{prana2018categorizing} manually coded 393 \emph{README.md} files and built a machine learning model that predicts the category a certain text would fall under, such as which part of the guidelines refers to who, what and why of the contribution process.
They do not consider if these guidelines are followed nor do they provide details on the contribution process itself. 

\subsection{Continuous Integration Tools}

CI tools offer a way to run automated checks on contributions that get submitted to software repositories, and Vasilescu et al.~\cite{vasilescu2015quality} show they increase contribution review efficiency.
Fowler and Foemmel \cite{fowler2000continuous} (and later Fowler and Humble \cite{fowler2017continuous}) define the functions of a CI tool as follows:
\begin{itemize}
    \item It should initiate an \emph{automated} build once a new change has been pushed to the shared mainline.
    \item It should assemble all required dependencies to build the project on the latest version of the shared mainline.
    \item It should build the latest version on the shared mainline.
    \item It should run the tests specified by developers on the latest version of the shared mainline.
    \item It should report the build results to developers.
\end{itemize}
Because of the benefits of using CI tools~\cite{vasilescu2015quality}, GitHub now offers a native, fully integrated CI solution
~\cite{github2019actions}. 
Yet, other CI tools are also available, e.g., the popular TravisCI~\cite{github2017ci}

Due to the role CI plays in evaluating code contributions on GitHub, developers have started considering CI among their contribution evaluation criteria \cite{gousios2015work, gousios2016work}.
Reviewers consider build results when reviewing code contributions, while contributors use them to evaluate their own contributions before submitting them.
It is, however, unclear how CI tools are discussed in contribution guidelines.
Thus, we focus on investigating the structure and contents of contribution guidelines, as well as how CI tools are featured in them.

\section{Methodology}
\label{sec:mthds}

For our investigation of GitHub project development practices and how they make use of continuous integration (CI) tools, 
we selected a cohort of GitHub projects from the GHTorrent dataset \cite{gousios2013ghtorrent}. 
We coded their contribution guidelines, as those generally offer documentation about contribution practices and the expectations core developers have about contributions.
This allowed us to answer RQ1 and RQ2, as well as determine the contents of the projects' contribution guidelines.
We also visualized the projects' activities on GitHub to observe their contribution processes and determine what type of development practices they follow.
This allowed us to answer RQ3 and explore the extent to which developers adhere to the prescribed practices.

\subsection{Project Selection Criteria}

In order to filter the large dataset provided by GHTorrent (about 37 million projects), we followed criteria laid out by Vasilescu et al.~\cite{vasilescu2015quality}, Tsay et al.~\cite{tsay2014influence}, and Munaiah et al.~\cite{munaiah2017curating}.
The combination of the criteria from the previously mentioned literature resulted in the following filters:
\begin{itemize}
	\item \textbf{Exclude forks:} Forks are typically created by a contributor who wishes to use a copy of the project's source code to make a contribution.
	Excluding them eliminates duplicates as well as incomplete project histories, as indicated by Tsay et al.~\cite{tsay2014influence} and Kalliamvakou et al.~\cite{kalliamvakou2014promises}.
	\item \textbf{Exclude deleted projects:} Deleted GitHub projects 
	are no longer accessible via the GitHub API and have been inactive for some time.
	Moreover, according to Kalliamvakou et al.~\cite{kalliamvakou2014promises}, their activity is deleted.
	\item \textbf{Exclude projects with no recent commits:} Commits indicate that a project is active and open to contribution.
	We considered projects that have at least one commit the week before the sampling period~\cite{kalliamvakou2014promises, tsay2014influence}.
	\item \textbf{Exclude projects that have less than 10 recent pull requests:} Pull requests, be they open or closed, represent contributions to a project, and thus represent project activity, as indicated by Gousios et al.~\cite{gousios2015work} and Vasilescu et al.~\cite{vasilescu2015quality}.
    We focused on projects where a contributor---particularly one who has no write privileges to the source repository---has access to the build results.
	\item \textbf{Exclude projects that have less than three unique contributors:} This is an indicator of the project having a tightly-knit community of developers that are actively collaborating but are less inclined to accept external contribution, as discussed by Munaiah et al. \cite{munaiah2017curating}.
	\item \textbf{Exclude projects that do not have at least one recently merged pull request:} According to Kalliamvakou et al. \cite{kalliamvakou2014promises}, having a pull request does not indicate that it was merged. 
	This criterion focuses on recently merged pull requests as a sign of a project \emph{accepting} contributions.
\end{itemize}

We determined how recent a commit or pull request was by whether or not it occurred in the week prior to the sampling phase.
The above combined criteria reduced the population to 41,642 projects that are non-duplicates, active, accept pull requests from contributors, and have a community of developers (or at least a team) supporting them.

The next step was to determine which projects use a CI tool.
We cloned the 41,642 projects that resulted from applying the previous filters to GHTorrent and mined their repositories for common CI tool configuration files (e.g., $.travis.yml$).
Based on this, the repositories were divided into two sets: those that use a CI tool (28,904 projects), and those that may not (12,738 projects).
While we followed the process outlined by Zampetti et al.~\cite{zampetti2017open}, we do note that some repositories may not have included a CI tool configuration file yet still use a CI tool.

The previously listed criteria, however, do not \emph{guarantee} the selection of a reasonably active project with a reasonably large community to accommodate the amount of activity we need for exploratory analysis.
To address this, we used GitHub's method of ranking open source repositories\footnote{\url{https://octoverse.github.com/projects#repositories}} by contributors.
We sorted the set of projects that use CI by the number of unique contributors and selected the top 100 projects.

For the most active projects that use CI tools, we coded their contribution guidelines.
We looked for a \emph{CONTRIBUTING.md} file first, and if that was not available, we then looked for a \emph{README.md} file.
We used those files as proxies for process documentation.
We excluded 28 of these 100 projects based on the following criteria:
\begin{itemize}
    \item The guideline file for a project is too small; less than 2 KB of data, similar to the filtering criteria used by Prana et al. \cite{prana2018categorizing}.
    \item The project guideline file contains no actual guidelines, rather it is only a link to an external source (typically style guides for particular languages)\footnote{Also similarly to Prana et al. \cite{prana2018categorizing}, we chose to only focus on files that GitHub initializes automatically. While it is possible that some may refer to an external source, these are usually much less common.}.
\end{itemize}
This left us with a final sample of 72 projects with high contribution activity that use CI tools and have substantive guidelines within their GitHub repositories.  

\subsection{Guideline Coding}

In order to understand how project team members envision their contribution processes, we examined their contribution guidelines (\emph{CONTRIBUTING.md}).
If the file did not exist in the repository, we inspected the project's basic documentation instead (\emph{README.md}).
We used thematic coding described by Creswell \cite{creswell2017research} in an inductive fashion to allow themes to emerge naturally.
For each of the 72 projects in our remaining sample, we went through their contribution guidelines, manually labeling every statement based on the topic it addressed.
For instance, ``\emph{If the code change needs to be applied to other branches as well (for example a bugfix needing to be backported to a previous version), one of the team members will either ask you to submit a PR with the same commit to the old branch, or do this for you.}'' was assigned to the ``How to Submit Bugfixes'' category.
And ``\emph{Please sign our Contributor License Agreement (CLA) before sending PRs. We cannot accept code without this.}'' fell under the ``Signing a CLA'' category.
As such, we constructed a coding index that grew with each file until we reached saturation after 50 files (we coded all 72 files, yet no additional codes emerged in our coding index). The full index is available as part of our reproducibility package\footnote{\url{https://figshare.com/s/c0d3321053380840d8fa}}.

Additionally, we compared our list of identified codes to those observed by Prana et al. \cite{prana2018categorizing} when they performed a similar activity (labeling README file contents for content classification via machine learning), as well as to the contribution process information gathered by Gousios et al. \cite{gousios2015work, gousios2016work} when they surveyed GitHub reviewers and contributors regarding their reviewing and contributing practices.
The codes we found were of a finer grain than those found by Prana et al. \cite{prana2018categorizing}, and as such, we were able to fit our codes into their higher-level categories.
Our codes also aligned with the results reported by Gousios et al. \cite{gousios2015work, gousios2016work} concerning pull request contributions.

\subsection{Project Workflow Mining and Visualization}

In order to better grasp a project's workflow in a way that accurately reflects the reality of the process as opposed to the documented version of the process, we mined the data from the GitHub events API. Unfortunately, only 53/72 projects were accessible via the API. 
We mined these 53 projects over a period of four weeks because inspecting the project workflows after that point showed little to no variation in terms of how a project processes contributions.
Over that period, we queried each projects' events API for events that happened throughout this period.
Such events included, but were not limited to:
\begin{itemize}
    \item opening/closing an issue;
    \item opening/closing a pull request;
    \item pushing a commit; and
    \item commenting on an issue/pull request/commit.
\end{itemize}

To get a better sense of each project's contribution process and determine if it matched the workflow prescribed in their contribution guidelines, we visually represented it as a process map.
We connected the various entities (issues, pull requests, commits, etc.) within the event logs already harvested to form a string of consecutive actions.
Where possible, we connected commits to their corresponding pull requests and pull requests to their corresponding issues based on the references developers made in the documentation of each artifact.

\begin{figure}
    \centering
    \includegraphics[width=0.87\columnwidth]{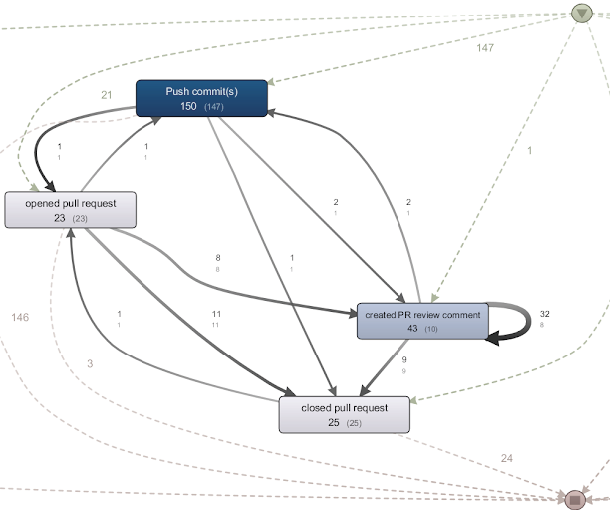}
    \vspace{-4mm}
    \caption{Excerpt from the Apache Camel process map.}
    \label{fig:petri_net}
    \vspace{-5mm}
\end{figure}

To visualize the contribution process for each project, we used the process mining tool disco\footnote{\url{https://fluxicon.com/disco/}}, 
which constructs process maps out of process logs to facilitate analysis. 
An example of the various paths a contribution can take is shown in Fig.~\ref{fig:petri_net}.
For instance, a contribution can be in the form of a commit directly made to the master branch, as illustrated by the push commit(s) step.
Some commits are also included as part of a pull request and elicit a code review. 
Alternatively, a commit can be made to a pull request, which then results in the pull request's closure.
Similarly, reviews can also result in the closure of a pull request.
%

\section{Results}
\label{sec:rslts}

Based on the methods we described, we were able to discern the contents of a typical contribution guideline file.
We also compared the prescribed contribution process to the actual process for the 53 projects of which we could mine the event API and that had substantive guideline documents.

\subsection*{RQ1: What is the content of contribution guidelines for projects on GitHub?}

Contribution guidelines embody the expectations integrators have for contributions to their projects.
We found five main categories of contribution guidelines: Project Orientation, Contribution Workflow, Pull Request Acceptance Criteria, Continuous Integration Tools, and Traceability.

The first category includes guidelines to introduce newcomers to a project and familiarize them with internal processes and workflows. Example sub-categories are details on how to submit issues and what sort of documentation is sufficient.
The second category, Contribution Workflow, typically walks contributors through the process of successfully submitting a pull request to a project.
Examples include how and when to create a new branch, how to create a pull request, and whether a Contributor License Agreement needs to be signed.
Under the category of Pull Request Acceptance Criteria we include statements that describe what reviewers consider to be an ideal pull request, using criteria such as contribution size, testability, and documentation.
The Continuous Integration Tools category includes themes about the usage of CI tools within the project's contribution process.
And finally, the Traceability category encompasses the theme of linking contribution process artifacts to each other.
Table~\ref{tbl:doc_cats} illustrates some of the most common themes across our sample.

\begin{table}[ht]
	\vspace{-4mm}
	\caption{Example of documentation category frequency} \footnotemark
    \label{tbl:doc_cats}
    \vspace{-2mm}
	\centering
	\smaller
	\begin{tabular}{l c}
		\toprule
		Content Category & Featuring Projects \\
		\midrule
		\emph{Pull Request Acceptance Criteria} & \\
		Contribution Style & 72.22\% \\
		Contribution Includes Test Cases & 52.78\% \\
		Contribution Documentation & 47.22\% \\
		\emph{Project Orientation} & \\
		How to Open an Issue & 69.44\% \\
		How to Set up a Local Development Environment & 48.61\% \\
		General Technical Knowledge & 38.89\% \\
		\emph{Contribution Workflow} & \\
		Submitting a Pull Request & 73.61\% \\
		How to Branch in a Repository & 56.94\% \\
		How to Fork/Clone a Repository & 52.78\% \\
		\emph{Continuous Integration Tools} & \\
		Testing by CI Tool & 30.56\% \\
		\emph{Traceability} & \\
		Artifact Linking for Traceability & 19.44\% \\
		\bottomrule
	\end{tabular}
	\vspace{-5mm}
\end{table}

\footnotetext{This table does not contain all coded themes. The full list can be found in the reproducibility package at \url{https://figshare.com/s/c0d3321053380840d8fa}.}

\subsection*{RQ2: Do projects that use CI tools mention these tools in their contribution guidelines?}

Based on the results we discussed above in Table \ref{tbl:doc_cats}, we found that CI tools were mentioned in only 31\% of our sample of contribution guideline documents.
When mentioned, it was only as a vehicle for running and passing tests as part of submitting a contribution.
There was no indication in the contribution guidelines as to whether a project followed the CI practice in terms of development workflow.
There was also no documentation regarding what these tools actually do or the scripts they run, compared to the dense amount of documentation that we found on other topics, including how to set up a development environment, and project structure.

\subsection*{RQ3: To what extent do the actual processes in projects that use CI tools match the processes in their guidelines?}

With respect to the contribution process workflow, we found that the actual activity trace data of the projects in our sample differed from the guidelines in the following ways:
\begin{itemize}
    \item Some projects made use of contribution practices that were not documented in the contribution guidelines, e.g.,
     51\% of the projects in our sample reopen issues, and 68\% reopen pull requests.
    However, the contribution guidelines offered no guidance on when or why a developer should reopen a previously closed issue or pull request.
    \item Fourteen projects (19.5\%) prescribed linking artifacts to each other for traceability reasons (see Table \ref{tbl:doc_cats}), yet we rarely observed occurrences of this happening.
    \item Although about 68\% of the projects whose activity we had access to described their contribution process in the form of creating and submitting pull requests, the contribution activity of all but one (i.e., 52/53 projects) involved direct commits to the master branch that were not linked to pull requests.
    Across all 53 projects, we found that the mean number of direct commits is 93\%, with a standard deviation of 11\% and a median of 99\%.
\end{itemize}
\section{Discussion}
\label{sec:discn}

Contribution guidelines are meant to be the first point of contact for developers who want to learn about the process a project team uses for development \cite{github2019contributing}.
They are designed to guide new developers and orient them around the project, telling them about the tools they need in order to make contributions effectively and efficiently.
However, our study of 53 active GitHub projects that use CI (and that we could mine) shows two major shortcomings in contribution guidelines: they do not accurately reflect all the agreed-upon methods of contribution, and they focus more on automatable details that a tool can check for than they do on the specifics of how to contribute.
The overwhelming majority (72\%) of projects we studied include guidelines about code style and other technical information.
Most of these details are automatable: code style, for example, can be efficiently checked with linters like Checkstyle.
This document real estate could be better used to surface and make explicit the tacit knowledge that core team members have about their processes and internal workflows.

Steinmacher et al. \cite{steinmacher2015systematic} suggest this tacit knowledge is more useful, as they found that a lack of knowledge regarding project components and processes is one of the barriers faced by newcomers.
This barrier could be alleviated by contribution guidelines that contain information on the \emph{contribution workflow}.
For example, we noticed a lack of CI tool documentation except for how to run the CI tool---there was no information on how the CI tool fits within the project's workflow.
While some projects include detailed information on the project's structure, dependencies, and the process one should follow in order to contribute effectively, several projects in our sample do not include adequate information.
About a quarter (26.4\%) of the sample projects do not prescribe workflow guidelines at all, and do not include any information on submitting pull requests or developer branching conventions.

Our future research will focus on the ways in which guideline documents, such as README files \cite{prana2018categorizing}, can assist new developers.
In particular, it is not clear to what extent the mandatory use of CI tools improves the process of contributing code to a new project.
We need to understand why contribution guidelines exist in the form they do now, and whether contributors consider them adequate sources of information.
We also need to explore why core team members do not adhere to the contributions they prescribe.
\section{Threats to Validity}

The limitations from our work include generalizability, in that we were limited to mining the workflow data from only 53 projects of the candidate 72 projects we considered in this research.
Our coding process may also be subject to bias, which we mitigated by referencing previous work on contribution guidelines ~\cite{prana2018categorizing}.

Our interpretation of the actual workflow process also relies on the Disco mining tool we used, however, we manually checked the results it produced.
We also use the contribution guidelines as a proxy for contribution process documentation, which should apply to both core team members as well as external contributors.
However, this is not always the case \cite{avelino2019measuring}.
Finally, it is possible that some projects define their contribution guidelines in other resources, but we tried to address this by following a similar process by Prana et al.~\cite{prana2018categorizing} to exclude these projects in our analysis.
\section{Conclusion}
\label{sec:confw}

Contribution guidelines embody a software project's contribution process, however, there has yet to be an exploration of what they contain and whether projects adhere to the workflows they prescribe.
We demonstrate that the most active projects that use CI in fact do not follow their own guidelines (if they have any) by conducting a mixed-methods study of these 53 GitHub projects using thematic coding of guideline documents and process mining of GitHub event streams. 
Furthermore, we speculate that the current contribution guideline structure may be written to suit project maintainers more than new contributors.
A more in-depth study of both process documentation and developer perceptions is needed in order to determine how effective the current guideline format is and whether it needs to be optimized for the contributor.

\section*{Acknowledgement}
This research is supported by the Natural Sciences and Engineering Research Council of Canada (NSERC).
We thank Cassandra Petrachenko for her help with this study.

\bibliographystyle{IEEEtran}
\bibliography{main.bib}

\end{document}